\documentclass[usegraphicx,usenatbib,onecolumn]{mn2e}

\begin{document}
\title{Probing Primordial Magnetic Fields with 
the 21cm Fluctuations}

\author[Tashiro, H. et al.]
{Hiroyuki Tashiro$^1$\
 and Naoshi Sugiyama$^2$\\
  $^1$Department of Physics, Kyoto University, Kyoto 606-8502, Japan\\
  $^2$Department of Physics and Astrophysics, Nagoya University, 
  Chikusa, Nagoya 464-8602, Japan}

\date{\today}

\maketitle

\begin{abstract}

Primordial magnetic fields possibly generated in the very early
universe are one of the candidates for the origin of magnetic fields
observed in many galaxies and galaxy clusters.  After recombination,
the dissipation process of the primordial magnetic fields increases
the baryon temperature.  The Lorentz force acts on the residual ions
and electrons to generate density fluctuations.  These effects are
imprinted on the cosmic microwave background (CMB) brightness
temperature fluctuations produced by the neutral hydrogen 21cm line.
We calculate the angular power spectrum of brightness temperature
fluctuations for the model with the primordial magnetic fields of a
several nano Gauss strength and a power-law spectrum.  It is found
that the overall amplitude and the shape of the brightness temperature
fluctuations depend on the strength and the spectral index of the
primordial magnetic fields.  Therefore, it is expected that the
observations of the CMB brightness temperature fluctuations give us a
strong constraint on the primordial magnetic fields.

\end{abstract}


\maketitle

\section{introduction}

Observations reveal existence of magnetic fields on very large
scale, i.e., galaxies and clusters of galaxies.  It is found that
these magnetic fields typically have a few $\mu $Gauss strengths and
relatively large coherent scales, i.e., a few tens of kpc for clusters
of galaxies and a few kpc for galaxies~\citep{magobs}.  The origin of
such magnetic fields has not yet known while many ideas have been
proposed.  Perhaps it may be one of the most important remaining
problems of cosmology and astrophysics to find out the origin and
evolution of magnetic fields in the history of the universe.

The most conventional idea is that such magnetic fields were formed
due to the astrophysical processes such as Biermann
battery~\citep{biermann} in stars and supernova explosions.  Then
these seed magnetic fields were amplified by the dynamo process.
Eventually, supernova winds or active galactic nuclei (AGN) jets may
spread these magnetic fields into inter-galactic medium (for a
comprehensive review see \citealt{magreview}).  However, it is still
little known about the efficiency of the dynamo process in the
expanding universe.  It is particularly difficult for the
astrophysical processes to explain observed magnetic fields with very
large coherent scales in galaxy
clusters~\citep{clustermag1,clustermag2}.  A recent observation
suggests existence of magnetic fields in high redshift
galaxies~\citep{maghighz}.  These galaxies may be dynamically too
young for the dynamo process to play a role.

An alternative scenario is that magnetic fields were formed in the
very early universe.  Many authors have suggested various generation
mechanisms of the primordial magnetic fields in the early universe.
One can introduce an exotic coupling between electro-magnetic and
scalar fields to generate magnetic fields during the inflation
epoch. Or one can consider bubble collisions during cosmological phase
transitions such as QCD or electroweak for generation of magnetic
fields. For a detailed review, see~\citet{giova}.
In this alternative scenario, one may directly obtain the nano Gauss 
primordial magnetic fields which are sufficient enough to explain 
$\mu$Gauss magnetic fields observed at present since 
the adiabatic compression due to the structure formation could easily
amplify the primordial magnetic fields by a factor $\sim 10^3$.  
In this case, there is no need of the dynamo process.  However, if the
seed magnetic fields generated in the early universe were too weak, 
the dynamo process is required even in this scenario while the
coherent length could be very large unlike the astrophysical processes.  

If the magnetic fields were generated in the very early universe, such
primordial magnetic fields may play an important role for various 
cosmological phenomena.  There are many previous works to constrain 
the strength of the primordial magnetic
fields from Big Bang Nucleosynthesis (BBN), temperature anisotropies and
polarization of Cosmic Microwave Background (CMB), or structure
formation.  These constraints 
give us clues to the origin of large-scale magnetic
fields, and when and how magnetic fields were generated.

The primordial magnetic fields affect on BBN through 
the enhancement of the cosmological expansion rate as neutrinos 
or the modification of the reaction rates of the light elements.
The limit on the magnetic field strength from BBN
is $B_0 \la 7 \times 10^{-5}$Gauss 
where $B_0$ is the comoving magnetic field 
strength~\citep{bbnconstraint1,bbnconstraint2}.

The black-body energy spectrum of CMB could be distorted by the
primordial magnetic fields through the dissipation process.  The
severe constraint on the spectrum distortion from COBE/FIRAS
observation leads to $B_0 \la 3 \times 10^{-8}$Gauss on comoving
scales of $400$--$600$pc at redshift $z \ga 2 \times
10^6$~\citep{diss-j-k-o}.

The primordial magnetic fields produce CMB temperature anisotropies on
various scales.  On very large scales, coherent magnetic fields
generate the anisotropic expansion of the universe.  Magnetic fields
with the present-horizon-size coherent scale are constrained as $B_0
\la 10^{-9}$Gauss~\citep{barrow-ferreira}.  On intermediate or small
scales, magnetic pressure modifies the acoustic oscillations of
baryon-photon fluids. Or Alfven mode (fluid vorticity) induces
temperature anisotropies.  Constraints on the primordial magnetic
fields with $100$Mpc--$1$Mpc coherent scales from WMAP and other CMB
experiments are $B_0 \la 10^{-8}$Gauss
\citep{adams-danielsson,mack,lewis,yamazaki,tashiro}.  Moreover CMB
polarization suffers from magnetic fields.  The Faraday rotation 
of the polarization is caused when CMB crosses 
the ionized inter-galactic medium (IGM) if there exist magnetic
fields~\citep{scoccola-harari-fara1,kosowsky-kahniashvili-fara2}.  
The fluid vorticity induced by magnetic fields can generate
B-mode (parity odd) polarization as well as E-mode (parity
even)~\citep{subramanian-seshadri-barrow,tashiro}.  We expect to have
much stringent limits on the primordial magnetic fields by future CMB
observations, e.g., {\it Planck}~(http://www.rssd.esa.int/Planck).

The primordial magnetic fields may play an important role for the
structure formation in the universe.  The Lorentz force of the
primordial magnetic fields could induce density fluctuations once the
universe becomes transparent after
recombination~\citep{wasserman,kim-olinto,subramanian-barrow-nonline,gopal-sethi}.
The magnetic tension and pressure are more effective on small scales
where the entanglements of magnetic fields are larger.  Therefore, if
there exist the primordial magnetic fields, it is expected that there
is the additional power in the density power spectrum on small scales
and these power induces the early structure
formation~\citep{sethi-subramanian,tashiro-reion}.

Thermal evolution of baryons after decoupling from photons, redshift
$z\sim 200$, could be also modified by the existence of the primordial
magnetic fields~\citep{sethi-subramanian} since the dissipation of the
primordial magnetic fields work as the heat source.  The dissipation
mechanisms are the ambipolar diffusion and the direct cascade decay of
magnetic fields.

Both the structure formation and the thermal evolution of baryons are
closely connected to reionization of IGM.  Hence, we can suspect
that the reionization process can be strongly affected by the
existence of the primordial magnetic fields.

One of the best probes of the reionization process of IGM is the CMB
brightness temperature fluctuations induced by neutral hydrogens
through the 21cm line.  In this paper, therefore, we examine the
effect of the primordial magnetic fields on the CMB brightness
temperature fluctuations produced by the redshifted hydrogen 21cm
line.  The hydrogen 21cm line is caused by a spin flip of the electron
in a neutral hydrogen atom.  The neutral hydrogen atoms produce the
brightness temperature fluctuation of CMB by 21cm line absorption from
and emission into CMB.  The amplitude of the brightness temperature
fluctuations depends on the hydrogen density, ionization fraction of
hydrogens and hydrogen temperature.  Therefore observations of the
brightness temperature fluctuations at wave length $21 (1+z)$cm
reveal the density
fluctuations and the ionization process at redshift $z$
\citep{loeb-zaldarriaga,bharadwaj-ali,cooray,ali-bharadwa-reion}.
Recent efforts to measure redshifted 21cm line such as 
LOFAR~(http://www.lofar.org),
MWA~(http://web.haystack.mit.edu/arrays/MWA/MWA.html) and
SKA~(http://www.skatelescope.org), will soon give us clues of dark
ages of the universe.  


This paper is organized as follows.  In Sec.~II, we discuss the effect
of the primordial magnetic fields on the hydrogen temperature and the
density fluctuations.  In Sec.~III, we summarize CMB bright
temperature fluctuations produced by the hydrogen 21cm line.  In
Sec.~IV, we compute the angular power spectrum of bright temperature
fluctuations with the primordial magnetic fields and discuss the effect of
the primordial magnetic fields.  Sec.~V is devoted to summary.  Throughout
the paper, we take WMAP values for the cosmological parameters, i.e.,
$h=0.71 \ (H_0=h \times 100 {\rm Km/s \cdot Mpc})$, $T_0 = 2.725$K,
$h^2 \Omega _{\rm b} =0.0224$ and $h^2 \Omega_{\rm m} =0.135$
\citep{spergel}.  And $\hbar$ and $c$ are Planck's constant over $2
\pi$ and speed of light, respectively.

\section{Effects of the primordial magnetic fields on thermal history
  and structure formation}

After recombination, the primordial magnetic fields affect the thermal
history and the structure formation of the universe.  In this section
we summarize how the existence of magnetic fields modifies them. 

Let us discuss the evolution of the primordial magnetic fields and the
resultant power spectrum.  First, we postulate that the primordial
magnetic field lines are frozen-in to baryon fluid in the early
universe.  This assumption leads to the time evolution as
\begin{equation}
{\bf B} _0({\rm x}) = a^2 (t) {\bf B}(t, {\rm x}),
\label{def-com}
\end{equation}
where $a(t)$ is the scale factor, which is normalized to the present
value, and ${\bf B}_0$ is the comoving strength of magnetic fields.
Since the baryon fluid is highly conductive through the entire history
of the universe, this relation can be hold as far as magnetic fields
do not suffer from non-linear processes such as direct cascade induced
by turbulent motion of eddies.

Next, we assume that the primordial magnetic fields are
isotropic and homogeneous Gaussian random fields, and have a
power-law spectrum as 
\begin{equation}
\langle B_{0i}({\bf k_1} ) B_{0j} ^* ({\bf k_2} )\rangle=
{(2 \pi)^3 \over 2 } \delta
({\bf k_1- k_2}) \left(\delta_{ij}-{k_{1i} k_{2i} \over k_1^2 } \right) B^2 _0 (k),
\label{isotropic}
\end{equation}
\begin{equation}
4 \pi k^3 B^2 _0 (k) = (n+3) \left({k \over k_{\rm c}} \right) ^{n+3} 
B_{\rm c} ^2 \qquad {\rm for}\ (k<k_{\rm c}),
\label{powerlaw}
\end{equation}
where $k_{\rm c}$ is the cutoff wave-number and $n$ is the spectral
index.  Here $B_{\rm c}$ is the root-mean-square amplitude of the magnetic field
strength in real space.

The cutoff in the power spectrum appears due to dissipation of 
magnetic fields associated with the nonlinear direct cascade
process~\citep{jedamzik-katalini,subramanian-barrow-nonline,banerjee-jedamzik}.
The direct cascade process transports the magnetic field energy from
large scales to small scales through the breaking of flow eddies and
generates the peak in the energy spectrum resultantly.  The time-scale
of the eddy breaking at the scale $l$ is $l/v$, where $v$ is the
baryon fluid velocity.  The direct cascade occurs when the eddy
breaking time-scale is equal to the Hubble time $H^{-1}$.  Once the
direct cascade takes place, the ``red'' power-law tail is produced in
the power spectrum of magnetic fields.  In order to simplify following
calculations, however, we assume that the direct cascade
process produces a sharp cutoff instead of the power-law cutoff in the spectrum.

Once the universe becomes transparent after recombination, baryons
decouple from photons and their velocity starts to increase.  Eventually the
velocity achieves the value determined by the equipartition
between the magnetic field energy and the kinetic energy of the baryon fluid,
namely the Alfven velocity, $v_{\rm A} \equiv c B_0 / \sqrt{4
\pi \rho_{{\rm b}0} a(t)}$ where $\rho_{{\rm b}}$ is the baryon
density and the subscript $0$ denotes the present value.  Accordingly,
the comoving cutoff scale induced by the direct cascade process is given by
\begin{equation}
k_{\rm c} \approx 2 \pi { H a\over v_{\rm A} } \approx 52{\rm
Mpc^{-1}} \left( {B_{\rm c} \over 1{\rm nGauss}} \right)^{-1} \left(
{h^2 \Omega_{\rm b}  \over 0.0224} \right)^{1/2}\left(
{h^2 \Omega_{\rm m} \over 0.135} \right)^{1/2}. 
\label{cutoff}
\end{equation}
Note that this comoving cutoff scale is time independent in the matter
dominated epoch.

\subsection{Dissipation of magnetic fields}

The energy of the primordial magnetic fields is
dissipated by the ambipolar diffusion and the direct cascade process.
The dissipation of magnetic field energy gives a considerable effect
to the thermal evolution of hydrogens.  If the energy of the
primordial magnetic fields with the strength $B_0$ is instantaneously converted
to the thermal energy of hydrogens at redshift $z$, 
the hydrogen temperature $T_{\rm k}$ can be estimated as $T_{\rm k}
\sim 30 (1+z) (B_0/1{\rm nGauss})^2$ K.

The ambipolar diffusion is caused by the velocity difference between
ionized and neutral particles.  Because magnetic fields accelerate
ionized particles by the Lorentz force while they give no effect on
neutral particles, the difference of velocity between ionized and
neutral particles arises.  This difference induces the viscosity of
ionized-neutral baryon fluid.  Accordingly, the energy of magnetic
fields is dissipated into the fluid.

On the other hand, the direct cascade is the nonlinear process 
which induces coupling between different modes.  When the time-scale of
the eddy breaking equals the Hubble time, the magnetic field energy is
transported from large scales to small scales by the nonlinear process
and the transported energy ends up dissipated.
The dissipation process due to the direct cascade, therefore, 
shifts the cutoff scale of the spectrum of the magnetic fields to
larger scales.  However it is shown in Eq.~(\ref{cutoff}) that 
the cutoff scale does not evolve during the matter dominated epoch.  
So we can conclude that the direct cascade is not the major source of
the dissipation of the magnetic fields. 
In fact, \citet{sethi-subramanian} have investigated the
effect of these dissipations on the cosmological thermal history and
have showed that the effect of the ambipolar diffusion dominates that
of the direct cascade process. 
In this paper, therefore,
we take into account only the ambipolar diffusion  as the dissipation
process.

The energy dissipation rate of the ambipolar diffusion is described
as~\citep{cowling}
\begin{equation}
\Gamma={  1 \over 16 \pi^2 \chi \rho_{\rm b}^2 x_{ e}} 
\left| \left(\nabla \times {\bf B}\right) \times {\bf B} \right|^2,
\label{rate-ambi-1}
\end{equation}
where $x_e$ is the ionization fraction and $\chi$ is the drag
coefficient for which we adopt the value computed by \citet{draine-roberge}
: $\chi = 3.5\times 10^{13} {\rm cm}^3 {\rm
g}^{-1} {\rm s}^{-1}$.  For the Gaussian statistics of the primordial
magnetic fields, Eq.~(\ref{isotropic}), the energy dissipation rate
can be rewritten as
\begin{equation}
\Gamma={ 7 \over 192 \pi^2 \chi \rho_{\rm b}^2 x_{\rm e} a^{10}}
\int dk_1 \int dk_2 B_0 ^2 (k_1) B_0 ^2(k_2) k_1 ^2 k_2 ^4.
\label{rate-ambi-2}
\end{equation}
The evolution of the hydrogen temperature with the
ambipolar diffusion, which is described by $\Gamma$,  is given by
\begin{equation}
{d T_{\rm k} \over d t }   =  -2 {\dot a \over a} T_{\rm k} + {x_e \over 1 + x_e} 
{8 \rho_\gamma \sigma_{\rm T} \over 3 m_e c } (T_{\gamma} - T_{\rm k}) 
+  { \Gamma  \over 1.5 k_{\rm B} n_e},
\label{evo-baryontemp} 
\end{equation}
where $\rho_\gamma$, $\sigma_{\rm T}$, $m_e$, $T_{\gamma}$ and $k_{\rm
B}$ are the CMB energy density, the Thomson cross-section, the
electron mass, the CMB temperature and the Boltzmann constant,
respectively.  The dot represents a derivative with respect to time.
For the standard cold dark matter dominated universe model, the
residual ionization fraction after recombination is $x_e \approx
10^{-4}$ until the universe reionizes.  However, it is possible that
higher hydrogen temperature due to the magnetic field dissipation
makes the collisional ionization effective and resultantly, the
ionization rate becomes higher.  The evolution of the ionization fraction
is described as~\citep{peebles1}
\begin{equation}
{d x_e \over dt} = \left[ \beta_e (1-x_e) 
-\alpha_e n_{\rm b} x_e^2 \right] C 
+ \gamma_e n_{\rm b} (1-x_e)x_e,
\label{evo-ionization}
\end{equation}
where 
\begin{equation}
\beta_e = {m_e k_B T_{\rm k} \over 2 \pi \hbar^2}^{3/2}
 \exp\left( - {\Delta E \over k_{ B} T_{\gamma}} \right),
\label{rate-ionization1}
\end{equation}
is the ionization rate out of the ground state with the ground state
binding energy $\Delta E=13.6$eV, $\alpha_e$ is the recombination rate to
excited states and $C$ is a suppression factor (for details, see
\citealt{peebles1,peebles2}).  The $\gamma _e$ in the second term is
the collisional ionization rate and we utilize the
fitting formula by \citet{Voronov}.  The collisional ionization
is suppressed by $\exp (- \Delta E/k_{\rm B} T_{\rm k})$.  Therefore, the
collisional ionization does not dominate the first term until the
hydrogen temperature exceeds $10^5$K.

\begin{figure}
  \begin{center}
    \includegraphics[keepaspectratio=true,height=50mm]{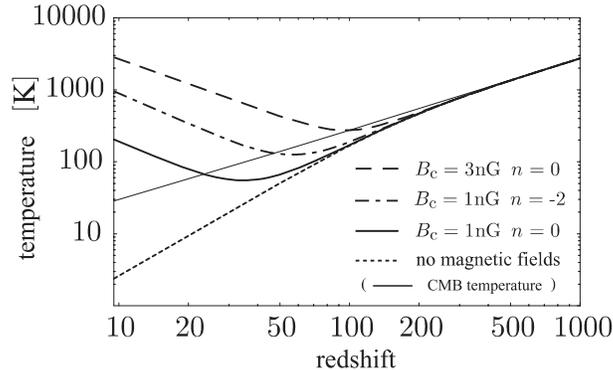}
  \end{center}
  \caption{The evolutions of the hydrogen temperature. 
  The dashed, dashed-dotted and solid lines are the hydrogen temperature 
  for magnetic fields with $B_{\rm c}=3$nGauss and $n=0$, $B_{\rm c}=1$nGauss and $n=-2$
  and $B_{\rm c}=1$nGauss and $n=0$.
  The dotted line is the hydrogen temperature in the standard cosmology
  (without magnetic fields) which is proportional to $(1+z)^{-2}$
  once hydrogens decoupled with photons at $z\sim 200$.
  The thin solid line represents the CMB temperature 
  which is proportional to $(1+z)^{-1}$.}
  \label{fig:temperature_redshift}
\end{figure}

We calculate Eqs.~(\ref{evo-baryontemp}) and (\ref{evo-ionization}) by
using the modified RECFAST code~\citep{recfast}.  We take into account
the ambipolar diffusion as the dissipation process of magnetic fields
and ignore other heat sources, e.g., heat from stars and galaxies.  We
plot the hydrogen thermal evolution in
Fig.~\ref{fig:temperature_redshift}.  In the redshift higher than
$z\sim 200$, the hydrogen temperature traces the CMB temperature due to
the second term in the right-hand side of Eq.~(\ref{evo-baryontemp}),
which represents the Compton scattering.  After $z\sim 200$ hydrogens
have decoupled from photons.  Since then, the hydrogen temperature
declines faster than the CMB temperature as $T_{\rm k} \propto
(1+z)^{2}$.  The heating by the dissipation of the magnetic field
energy gradually becomes effective and eventually the hydrogen
temperature goes up.  The heating efficiency of the dissipation
depends on not only the magnetic field strength but also the power-law
index of the spectrum.  This is because the magnetic fields with a
shallower spectrum produce the velocity difference between ionized and
neutral hydrogens on a wider range of scales.  Therefore, the
dissipation from magnetic fields with a smaller spectral index is more
effective than that with a larger spectral index since we normalize
the magnetic strength at the cutoff scale.

\begin{figure}
  \begin{center}
    \includegraphics[keepaspectratio=true,height=50mm]{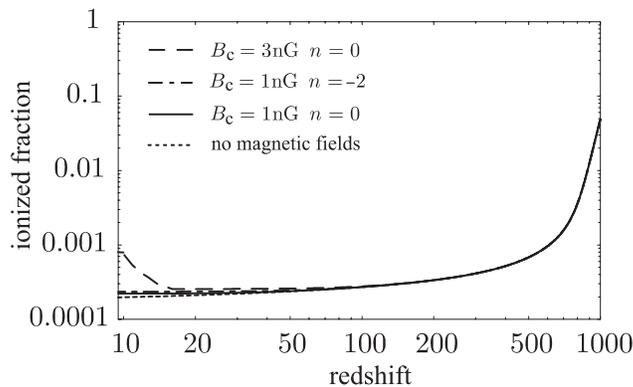}
  \end{center}
  \caption{The evolution of the ionization fraction.
  The dashed, dashed-dotted and solid lines are the ionization fraction 
  for magnetic fields with $B_{\rm c}=3$nGauss and $n=0$, $B_{\rm c}=1$nGauss and $n=-2$
  and $B_{\rm c}=1$nGauss and $n=0$.
  The dotted line is the ionization fraction in the standard cosmology
  (without magnetic fields).
  In the neutral hydrogen number density, 
  there is no significant change due to the energy diffusion of magnetic fields. }
  \label{fig:ionizedfraction_redshift}
\end{figure}

Fig.~\ref{fig:ionizedfraction_redshift} shows the evolution of the
ionization fraction.  If the hydrogen temperature is higher than
$5\times 10^5$K, the collisional ionization becomes effective (see
\citealt{sethi-subramanian}).  In our case, however, the hydrogen
temperature never exceeds this value so that the modification of the
ionization fraction by the magnetic field dissipation is very little.

\subsection{Generation of density fluctuations}

Primordial magnetic fields generate the density fluctuations after
recombination~\citep{wasserman,kim-olinto,subramanian-barrow-nonline,gopal-sethi,sethi-subramanian}.
The magnetic tension and pressure are more effective on small scales
because the entanglements of magnetic fields are larger.  Therefore,
on small scales, the additional density power by magnetic fields is
expected to dominate the primordial density power spectrum produced by
inflation.  The evolution equations of the density fluctuations with
the primordial magnetic fields are described as,
\begin{equation}
{\partial^2 \delta_{\rm b} \over \partial t^2} = -2 {\dot a \over
a}{\partial \delta_{\rm b} \over \partial t } +4 \pi G (\rho _{\rm b}
\delta_{\rm b} + \rho _{\rm dm} \delta_{\rm dm} ) + S(t,{\bf x}),
\label{baryon-den-eq}
\end{equation}
\begin{equation}
S(t,{\bf x})={ \nabla \cdot \left( (\nabla \times {\bf B_{0} ({\bf
x})}) \times {\bf B_{0} ({\bf x})}\right) \over 4 \pi \rho_{{\rm b} 0}
a^3 (t) },
\end{equation}
\begin{equation}
{\partial^2 \delta_{\rm dm} \over \partial t^2} = -2 {\dot a \over
a}{\partial \delta_{\rm dm} \over \partial t } +4 \pi G (\rho _{\rm b}
\delta_{\rm b} + \rho _{\rm dm} \delta_{\rm dm} ) ,
\label{dm-den-eq}
\end{equation}
where $\rho_{\rm dm}$ is the dark matter density, and 
$\delta_{\rm b}$ and $\delta_{\rm dm}$ are   
the density contrasts of baryons and dark matters, respectively.

In order to solve these equations, we define 
the total matter density $\rho_{\rm m}$ 
and the matter density contrast $\delta_{\rm m}$ as
\begin{equation}
\rho_{\rm m} \equiv \rho_{\rm b}+\rho_{\rm dm},
\label{rhom-defi}
\end{equation}
\begin{equation}
\delta_{\rm m} \equiv { (\rho _{\rm b} \delta_{\rm b} + \rho _{\rm dm} \delta_{\rm dm} ) \over \rho_{\rm m}}.
\label{deltam-defi}
\end{equation}
From Eqs.~(\ref{baryon-den-eq}) and (\ref{dm-den-eq}),
the evolution of $\delta _{\rm m}$
is given by
\begin{equation}
{\partial^2 \delta_{\rm m} \over \partial t^2} = -2 {\dot a \over
a}{\partial \delta_{\rm m} \over \partial t } +4 \pi G \rho _{\rm m}
\delta_{\rm m} + {\rho_{\rm b} \over \rho_{\rm m}} S(t,{\bf x}).
\label{matter-den-eq-2}
\end{equation}
The solution of Eq.~(\ref{matter-den-eq-2}) can be acquired
by the Green's function method,
\begin{eqnarray}
\delta_{\rm m} =A({\bf x}) D_1 (t)+ B({\bf x}) D_2 (t) 
-{\Omega_{\rm b} \over \Omega_{\rm m}} D_1(t) \int^{t}_{t_{\rm i}} 
dt' {S(t' ,{\bf x}) D_2 (t') \over W(t')}
+{\Omega_{\rm b} \over \Omega_{\rm m}} D_2(t) \int^{t}_{t_{\rm i}} 
dt' {S(t' ,{\bf x}) D_1 (t') \over W(t')},
\label{solution-matter-eq}
\end{eqnarray}
where $D_1(t)$ and $D_2(t)$ are the homogeneous solutions
of Eq.~(\ref{matter-den-eq-2})
and $W$ is the Wronskian and is expressed as
\begin{equation}
W(t)=D_1(t) \dot D_2(t) -D_2(t) \dot D_1(t),
\label{wronskian}
\end{equation}
and $t_{\rm i}$ denotes the initial time.

The first and second terms of Eq.~(\ref{solution-matter-eq})
correspond to the growing and the decaying mode solutions of
primordial density fluctuations and the third and fourth terms are the ones
generated by the primordial magnetic fields.  We represent the former
two terms as $\delta_{\rm mP}$ and the latter twos as $\delta_{\rm mM}$.
Here we only consider the growing solution for $\delta_{\rm mP}$.  The
analytic solution of $\delta_{\rm mM}$ in the matter dominated epoch
is obtained by \citet{wasserman} and \citet{kim-olinto}.  In this
paper, we concentrate on the evolution of fluctuations before
reionization $z \ga 10$ when the universe is matter dominant and the
contribution from the dark energy is negligible.  Therefore we can
utilize their solution.  In the matter dominated epoch, $D_1(t)
\propto t^{2/3}$ and $D_2(t) \propto t^{-1}$.  Accordingly the terms
generated by magnetic fields of Eq.~(\ref{solution-matter-eq}) can be
written as
\begin{equation}  
\delta_{\rm mM} = {\Omega _{\rm b} \over \Omega _{\rm m}}
\left[{9 \over 10} \left({t \over t _{\rm i} }\right)^{2/3} 
+{3 \over 5} \left({t \over t _{\rm i} }\right)^{-1} -{3 \over 2}\right] t_{\rm  i} ^2 S( t_{\rm i}, {\bf x}).
\label{mag-part}
\end{equation}
Once the density fluctuations were generated by the primordial
magnetic fields, they grow due to the gravitational instability so that
the growth rate is same as the primordial density
fluctuations.

Next, we calculate the power spectrum of the matter density
fluctuations.  Assuming that there is no correlations between the
magnetic fields and the primordial density fluctuations for simplicity, we
can describe the
matter power spectrum as
\begin{equation}
P_{\rm m}(k) = P_{\rm mP}(k)+P_{\rm mM}(k) 
\equiv \langle |\delta_{\rm mP}(k)|^2 \rangle + \langle |\delta_{\rm
  mM}(k)|^2 \rangle, 
\label{matter-power}
\end{equation}
where $\delta_{\rm mP}(k)$ and $\delta_{\rm mM}(k)$ are the Fourier
components of $\delta_{\rm mP}$ and $\delta_{\rm mM}$, respectively,
and $\langle ~ \rangle$ denotes the ensemble average.

We numerically calculate $P_{\rm mP}(k)$ by using the CMBFAST
code~\citep{cmbfast}, while $P_{\rm mM}(k)$ is obtained from
Eq.~(\ref{mag-part}) as
\begin{equation}
P_{\rm mM}(k) = \left ( {\Omega _{\rm b} \over \Omega _{\rm m}} \right
) ^2 \left ( t_{\rm i}^2 \over 4 \pi \rho_{{\rm b}0}a^3 (t_{\rm i})
\right)^2 \left[{9 \over 10} \left({t \over t _{\rm i} }\right)^{2/3}
+{3 \over 5} \left({t \over t _{\rm i} }\right)^{-1}-{3 \over
2}\right]^2 I^2 (k),
\label{power-mag-part}
\end{equation}
where 
\begin{equation}
I^2 (k) \equiv \langle |\nabla \cdot (\nabla \times {\bf B}_0 ({\bf
x})) \times {\bf B}_0 ({\bf x})|^2 \rangle.
\label{nonline-convo}
\end{equation}
Applying isotropic Gaussian statistics Eq~(\ref{isotropic}),
we can rewrite the nonlinear convolution Eq.~({\ref{nonline-convo}}) 
as 
\begin{equation}
I^2 (k) = \int dk_1 \int d \mu {B^2 _0 (k_1) B^2 _0 (|{\bf k - k_1}|)
\over |{\bf k - k_1}|^2 } \left[2 k^5 k_1^3 \mu+ k^4 k_1 ^4 (1-5 \mu^2) +
2k^3 k_1^5 \mu^3 \right],  
\label{nonline-convo-2}
\end{equation}
where $\mu$ is $\mu = {\bf k} \cdot {\bf k}_1/ k k_1$.
The integration of $I^2(k)$ is determined by the value of integrand at
$k_1=k_{\rm c}$ and $\vert{\bf k_1- k}\vert =k_{\rm c}$ because the
power spectrum $B_0^2(k)$ has the power law shape with sharp cutoff at
$k=k_{\rm c}$.  Note that the direct cascade process of the magnetic
fields from large scales to small scales likely produces a power-law
tail instead of sharp cutoff in the power spectrum above $k_{\rm c}$
as we discussed before.  However, it is still true that most of the
contribution on the integration $I^2(k)$ comes from the peak of the
spectrum at $k_{\rm c}$ as well as the sharp cutoff case, although
some corrections may be needed.

Let us introduce an important scale for the evolution of
density fluctuations, i.e., the magnetic Jeans length.  
Below this scale, the magnetic pressure gradients, which we do not
take into account in Eq.~(\ref{matter-den-eq-2}), 
counteract the gravitational force 
and prevent further evolution of density fluctuations. 
The magnetic Jeans scale $k_{\rm MJ} \sim v_{\rm A}/H$ is evaluated
as~\citep{kim-olinto} 
\begin{equation}
k_{\rm MJ}=5 \pi { \sqrt{\rho_{\rm m0}  \rho_{\rm b0} G} \over B_0}
= 12.7 {\rm Mpc}^{-1} \left({B_0 \over 1{\rm nGauss}}\right)^{-1} 
\left({h^2 \Omega_{\rm b} \over 0.0224}\right)^{1/2} \left({h^2 \Omega_{\rm m} \over 0.135}\right)^{1/2}.
\label{jeans}
\end{equation}
Here, the primordial magnetic fields have a power-low spectrum with
spectral index $n$ and sharp cutoff at $k_{\rm c}$ so that the
magnetic Jeans scale can be written from Eqs.~(\ref{cutoff})
and~(\ref{jeans}) as
\begin{equation}
k_{\rm MJ} = \left( {5 \over 8\pi}\sqrt{3 \over 2}\right)^{2/(n+5)} k_{\rm c}.
\end{equation}  
Baryon density fluctuations below the magnetic Jeans scale 
(the mode $k > k_{\rm MJ}$) oscillate and do not grow.

\begin{figure}
  \begin{center}
    \includegraphics[keepaspectratio=true,height=50mm]{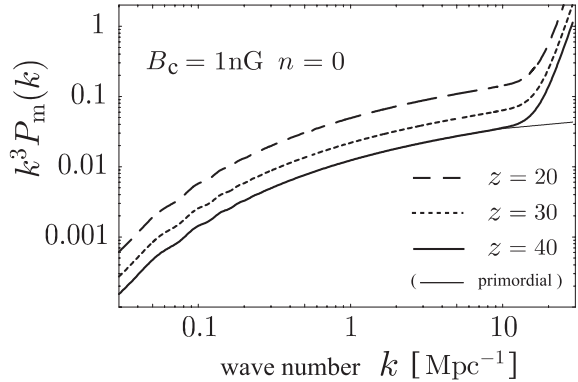}
  \end{center}
  \caption{The redshift evolution of the matter power spectrum for the model
  with $B_{\rm c}=1$nGauss and $n=0$.  The dashed, dotted and solid lines represent 
  the matter power spectrum at $z=20$, $z=30$ and $z=40$ (from top to bottom).
  The thin solid line is the matter power spectrum without
  magnetic fields at $z=40$.  
  }
  \label{fig:densitypert_redshift}
\end{figure}

Using Eq.~(\ref{nonline-convo-2}), we numerically calculate the matter
power spectrum $ P_{\rm m}(k)$.  We show the evolution of $k^3 P_{\rm 
m}(k)$ for the primordial magnetic fields with $B_{\rm c}=1$nGauss and $n=1$
in Fig.~\ref{fig:densitypert_redshift}.  The contribution from the
density fluctuations generated by the primordial magnetic fields is
dominated on small scales ($ \la 100$kpc).  Since the growth rates of
both $P_{\rm mP}$ and $P_{\rm mM}$ are $t^{4/3} \propto 1/(1+z)^2$,
the amplitude of the total matter power spectrum $P_{\rm m}$ is
proportional to $1/(1+z)^2$ and the comoving scale on which 
$P_{\rm mM}$ starts to dominate $P_{\rm mP}$ stays constant ($k \sim 15
{\rm Mpc}^{-1})$ as is shown in
Fig.~\ref{fig:densitypert_redshift}.

\begin{figure}
  \begin{center}
    \includegraphics[keepaspectratio=true,height=50mm]{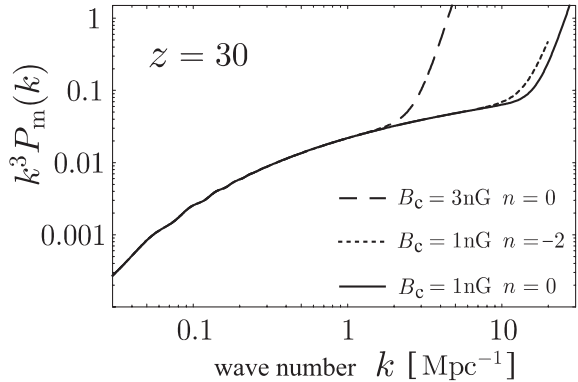}
  \end{center}
  \caption{The matter power spectra for models with different
  magnetic field strengths.  
   The dashed, dotted and solid lines are the power spectra 
   with the primordial magnetic fields $B_{\rm c}=3$nGauss and $n=0$, 
   $B_{\rm c}=1$nGauss and $n=-2$ and $B_{\rm c}=1$nGauss and $n=0$ (from top to
  bottom) at $z=30$.
   }
  \label{fig:densitypert_variosmag}
\end{figure}

In Fig.~\ref{fig:densitypert_variosmag}, we plot the matter power
spectra for models with different magnetic field amplitudes at $z=30$.
We can analytically estimate Eq.~(\ref{nonline-convo-2}) in the limit
of $k/k_ {\rm c} \ll 1$ as $I^2(k) \sim \alpha B_{\rm c}^{2n+10} k^{2n+7} +
\beta B_{\rm c}^7 k^{4}$ where $\alpha$ and $\beta$ are coefficients which
depend on $n$~\citep{kim-olinto, gopal-sethi}.  Here we employ the
fact that the cutoff scale $k_{\rm c}$ is proportional to $B^{-1} _{\rm c}$
as is shown in Eq.~(\ref{cutoff}).  The former term dominates if
$n<-1.5$, while the latter one dominates for $n>-1.5$.  Accordingly,
the matter power spectrum, $ k^3 P_{\rm m}(k) $, is proportional to $
B_{\rm c} ^{2n+10} k^{2n +7} $ for $n<-1.5$ or to $ B_{\rm c} ^7 k^{4} $ for
$n>-1.5$.  These dependences of $k^3P_{\rm m}(k)$ on $B_{\rm c}$ and $n$ can
be found in Fig.~\ref{fig:densitypert_variosmag}.

\section{brightness temperature fluctuations by the 21cm line}

In this section, we discuss the calculation of the CMB brightness
temperature fluctuations generated by the hydrogen 21cm line and the
angler power spectrum of them.  First, let us discuss the spin
temperature on which the amplitude of the brightness temperature
fluctuations depends.  The spin temperature is defined through the
ratio of the number density of hydrogen atoms between in the
excited state and in the ground state of the 21cm transition,
 \begin{equation}
{n_1 \over n_0 }= 3 \exp \left(-{ T_* \over T_{\rm s} } \right),
\label{def-spintemp}
\end{equation}
where subscripts 1 and 0 denote the excited and ground states and
$T_*$ corresponds to the energy difference between these states,
$T_*=0.0682$K.  The factor 3 in Eq.~(\ref{def-spintemp}) comes from
the ratio of the spin degeneracy factors.  The evolution of the number
density of hydrogen atoms in the ground state is described as
\begin{equation}
{\partial n_0 \over \partial z} - {3 \over 1+z} n_0 =-{1 \over (1+z)
H(z)} \left[ -n_0 \left( C_{01} + B_{01} I_\nu \right)+n_1 \left(
C_{10}+A_{10}+B_{10}I_\nu \right)\right],
\label{evo-ground}
\end{equation}
where $C_{01}$ and $C_{10}$ are the collisional excitation and
de-excitation rate coefficients, $A_{10}$ is the Einstein
A-coefficient, $B_{01}$ and $B_{10}$ are the Einstein B-coefficients
and $I_\nu$ is the specific intensity of CMB at frequency $\nu$.  The
collisional de-excitation coefficient is written as $C_{10}=4
\kappa_{10}(T_{\rm k}) n_{\rm H}/3$ where $n_{\rm H} \equiv n_0+n_1$
is the neutral hydrogen number density and $\kappa_{10}(T_{\rm k})$ is
tabulated as a function of the hydrogen temperature $T_{\rm k}$
\citep{allison-dalgarno}.  Typically $\kappa_{10}(T_{\rm k})$ is order
$10^{-10}{\rm cm}^3 {\rm s}^{-1}$ in $T_{\rm k} <1000$K.  The
excitation coefficient is related with the de-excitation coefficient
by the detailed balance as $C_{01}=C_{10} \exp(-2 \pi \hbar \nu /k
T_{\rm k})$.  
The Einstein coefficients are coupled by the Einstein relations 
as 
\begin{equation}
B_{01}=3 B_{10} = \left(3 c^2 \over 4 \pi \hbar \nu^3 \right) A_{10},
\label{einstein-relations}
\end{equation} 
where $A_{10}= 2.85 \times 10^{-15} {\rm s}^{-1}$.
The evolution of the number density of the excited hydrogen atoms
is also written as
\begin{equation}
{\partial n_1 \over \partial z} - {3 \over 1+z} n_1 =-{1 \over (1+z)
H(z)} \left[n_0 \left( C_{01} + B_{01} I_\nu \right) -n_1 \left(
C_{10}+A_{10}+B_{10} I_\nu \right)\right].
\label{evo-excited }
\end{equation}
From Eqs.~(\ref{def-spintemp}), (\ref{evo-ground}) and (\ref{evo-excited }),
we can derive the evolution of the spin temperature
\begin{equation}
{\partial \over \partial z} \left(1 \over T_{\rm s}\right)
=-{4  \over (1+z) H(z)  } \left[ \left({ 1 \over T_{\rm k} }-{ 1 \over T_{\rm s} } \right)C_{10}
+\left({ 1 \over T_{\gamma}} -{ 1 \over T_{\rm s} } \right){ T_{\gamma} \over T_{ *} }A_{10}\right],
\label{evo-spintemp}
\end{equation}
where we employ the approximation $\exp(-T_*/T)\approx 1-T_*/T$, and
the Rayleigh-Jeans law, $I_{\nu *}=2 \nu_* ^2 k_{\rm B} T_\gamma /c^2$
with $\nu_* \equiv k_{\rm B}T_*/(2\pi \hbar)$ because we are
interested in the epoch when $T_{\rm s}$, $T_{\rm k}$, $T_{ \gamma}
\gg T_{\rm *}$.

The emission and absorption of the 21cm line from neutral hydrogens 
affect the CMB brightness temperature at the frequency $\nu_* / (1+z)$ by
\begin{equation}
T_{\rm b} \left( {\nu_* \over (1+z)} ,{\bf\hat{n}} \right) = \tau { T_{\rm s }-T_\gamma  \over 1+z }, 
\label{brightnesstemp}
\end{equation}
where ${\bf\hat{n}}$ is a unit direction vector and $\tau$ is the optical depth of the 21cm line,  
\begin{equation}
\tau={3 \pi c^3 n_{\rm H} \hbar A_{10} \over 16 \nu_{*} ^2 k_{\rm B} T_{\rm S}}
{1 \over H(z)}
.
\label{opt-depth}
\end{equation}

The fluctuations of the brightness temperature at the frequency $\nu_*
/ (1+z)$ are obtained from Eq.~(\ref{brightnesstemp}) as
\begin{equation}
\delta T_{\rm b} \left( {\nu_* \over (1+z)} ,{\bf \hat{n}}\right) = 
{ \tau  \over 1+z } \left[ \left( T_{\rm s }-T_\gamma \right) \delta_{\rm m}
+T_\gamma {\delta T_{\rm s} \over T_{\rm s}}\right],
\label{pert-brightness}
\end{equation}
where we ignore the fluctuations of the CMB temperature and the
ionization fraction and we assume that the hydrogen number density
contrast is corresponding to the matter density contrast.  
From Eq.~(\ref{pert-brightness}), it is found that 
the brightness temperature fluctuations
at the frequency $\nu_* / (1+z)$ are determined by the density and the
spin temperature fluctuations at redshift $z$.  Note that we 
ignore the effect of the line-of-sight
component of the neutral hydrogen's peculiar velocity.  The
line-of-sight component of the peculiar velocity affects the optical
depth through the distortion of the redshift space.  The contribution
from the peculiar velocity could be substantially large on very large
scales (for details, see \citealt{bharadwaj-ali}), while it is
negligible on scales comparable to the future observations' survey
area.  Therefore we ignore the term of the peculiar velocity.

In order to calculate Eq.~(\ref{pert-brightness}), we need to know the
time evolution of the spin temperature fluctuations.  The evolution
equation can be obtained from Eq~(\ref{evo-spintemp}) as
\begin{equation}
{ \partial \over \partial z }\left({ \delta T_{\rm s} \over T_{\rm s}
}\right) ={ 4T_{\rm s} \over H(z)(1+z) } \left[\left({ C_{10} \over
T_{\rm k } }+{ A_{10} \over T_* }\right){ \delta T_{\rm s} \over
T_{\rm s} } +\left({ 1 \over T_{\rm k} }-{ 1 \over T_{\rm s}
}\right)C_{10}{ \delta n_{\rm H} \over n_{\rm H}} +\left(\left({ 1
\over T_{\rm k} }-{ 1 \over T_{\rm s} }\right){\partial \ln \kappa
\over \partial \ln T_{\rm k}} -{ 1 \over T_{\rm k } } \right)C_{10}{
\delta T_{\rm k} \over T_{\rm k} } \right],
\label{evo-deltaspin}
\end{equation}
where we ignore the fluctuations of the CMB temperature and the
ionization fraction.
Because of the energy pumping from CMB, strictly speaking, the
ordinary adiabatic relation, i.e. $\delta T_{\rm k}/T_{\rm
k}=(\gamma-1) \delta n_{\rm H}/n_{\rm H}$ where $\gamma$ is adiabatic
index and $\gamma=5/3$, is broken.  However, \citet{bharadwaj-ali} have
shown that the adiabatic relation is still valid until the star
formation process proceeds.  
Hence, hereafter, we assume $\delta T_{\rm k}/T_{\rm k}=(\gamma-1) \delta n_{\rm
H}/n_{\rm H}$ in Eq.~(\ref{evo-deltaspin}). 

Let us calculate the angular power spectrum of brightness temperature
fluctuations $C_l(\nu_* / (1+z))$.  We can expand the brightness
temperature fluctuations $\delta T_{\rm
b}\left(\nu_*/(1+z),{\bf\hat{n}}\right) $ into spherical harmonics
with the expansion coefficients $a_{lm}$.  
Accordingly the angular power spectrum $C_l(\nu_* / (1+z))= \left\langle
|a_{lm}|^2 \right\rangle$ can be written as 
\begin{equation}
C_l \left(\nu_* \over (1+z) \right)= \langle \vert
a_{lm}\vert^2\rangle =
4 \pi \int {dk^3 \over (2 \pi)^3} \left\langle |\delta T_{\rm b}
\left({\nu_* \over (1+z)}, k\right)|^2 
\right\rangle j_l ^2(k (\eta_0-\eta(z))),
\label{21cmangular-power}
\end{equation}
where $j_l (x)$ is the spherical bessel function, $\eta$ is the conformal time and $\delta T_{\rm
b}(\nu, k)$ is the Fourier component of $\delta T_{\rm b}$ at the
frequency $\nu$ which is calculated by Eq.~(\ref{pert-brightness}).

\section{results and discussion}

We can calculate the angular power spectrum of the brightness
temperature fluctuations produced by the 21cm line with the existence
of the primordial magnetic fields from Eq.~(\ref{21cmangular-power}).
In this section, we show the evolution of the spin temperature and
the angular power spectrum for three different primordial magnetic
fields, i.e., $(B_{\rm c},~ n)=(1{\rm nGauss},~0)$, 
$(1{\rm nGauss},~-2)$, and $(3{\rm nGauss},~0)$. 

\begin{figure}
  \begin{center}
    \includegraphics[keepaspectratio=true,height=120mm]{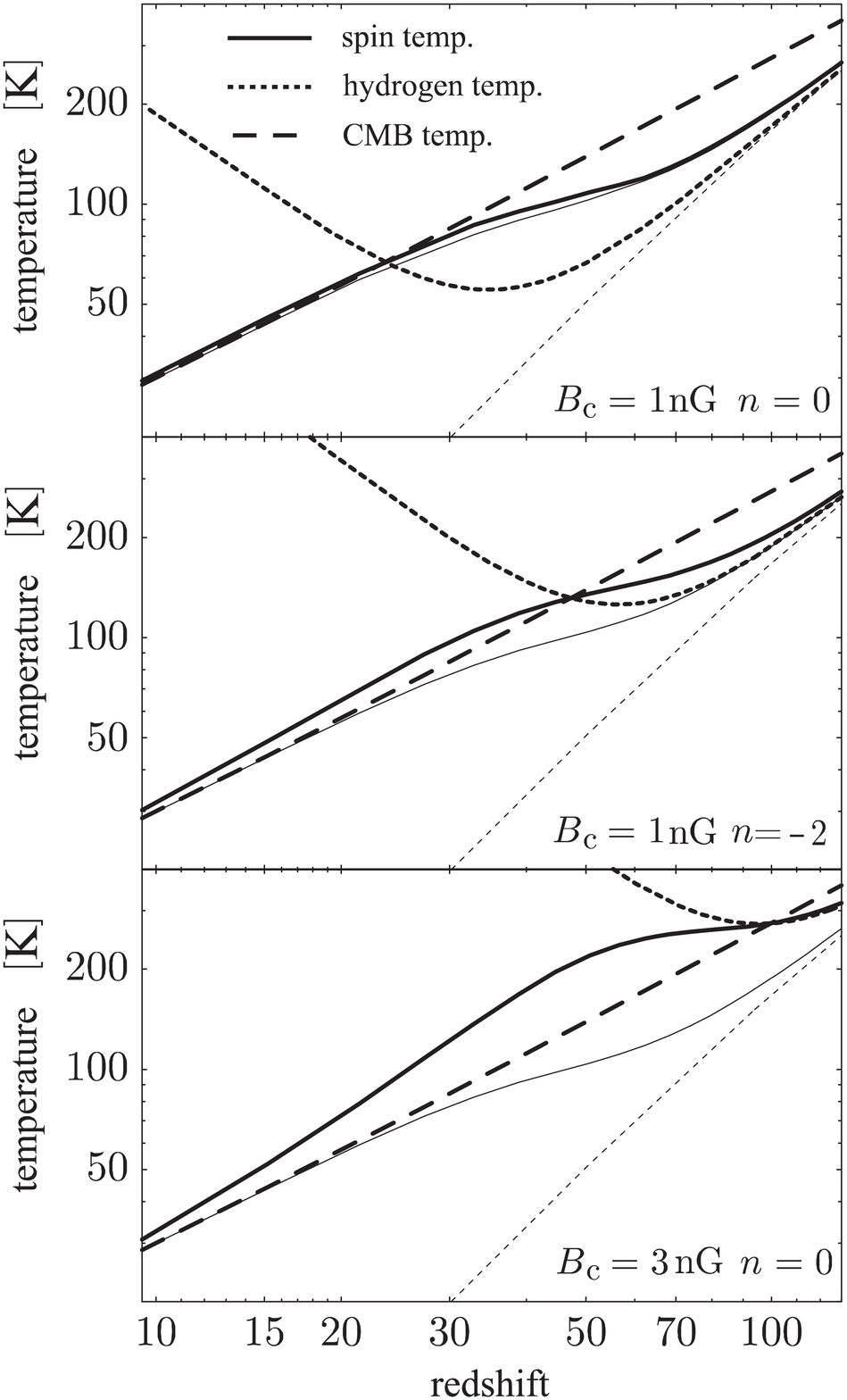}
  \end{center}
  \caption{The evolution of the spin temperature. 
  The solid lines, 
  the dotted and dashed lines denote the spin temperature, the hydrogen temperature and the CMB temperature, respectively.
  We show these temperature for three different magnetic field spectra:
  $B_{\rm c}=1$nGauss and $n=0$, $B_{\rm c}=1$nGauss and $n=-2$ and $B_{\rm c}=3$nGauss and $n=0$
  (from top panel to bottom panel).
  For reference, we plot the spin temperature and the hydrogen temperature in the standard cosmology
  with no primordial magnetic fields as the thin solid lines and dotted lines. 
  In all cases, the spin temperature closes in the hydrogen temperature in high redshift
  and approaches the CMB temperature asymptotically.}
  \label{fig:spintempfig}
\end{figure}

First we discuss the effect of the primordial magnetic fields on the spin
temperature.  Fig.~\ref{fig:spintempfig} shows the redshift evolution
of the spin temperature.  The spin temperature is well approximated to
the steady state of Eq.~(\ref{evo-spintemp}) as \citep{field}
\begin{equation}
T_{\rm s} = {T_\gamma + y T_{\rm k} \over 1+y},
\label{approx-spin}
\end{equation}
where $y$ is the collisional efficiency
and written as
\begin{equation}
y={2 \pi \hbar \nu_* C_{10} \over k_{\rm B} T_{\rm k} A_{10}}.
\label{collisonalefficiency}
\end{equation}
The behaviors of the spin temperature in Fig.~\ref{fig:spintempfig}
can be explained by Eq.~(\ref{approx-spin}).  In the high redshift
($z>100$), because the collisional efficiency $C_{10}$ is high ($y \gg
1$) so that the spin temperature couples the hydrogen temperature, i.e.,
$T_{\rm s} \simeq T_{\rm k}$.  As the universe expands the collisional
efficiency becomes low ($y \ll 1$) so that the spin temperature
approaches the CMB temperature $T_{\rm s} \simeq T_{\gamma}$.  The
primordial magnetic fields make the hydrogen temperature increase due to
the dissipation.  The collisional efficiency increases as the hydrogen
temperature increases.  Accordingly the combination $yT_{\rm K}$ in
the numerator of Eq.~(\ref{approx-spin}) increases although $y$
decreases.  Therefore the spin temperature becomes higher than the
one without the magnetic fields.  If the primordial magnetic fields
are strong enough, the hydrogen and spin temperatures can exceed the
CMB temperature as is shown in Fig.~\ref{fig:spintempfig}.

\begin{figure}
  \begin{center}
    \includegraphics[keepaspectratio=true,height=120mm]{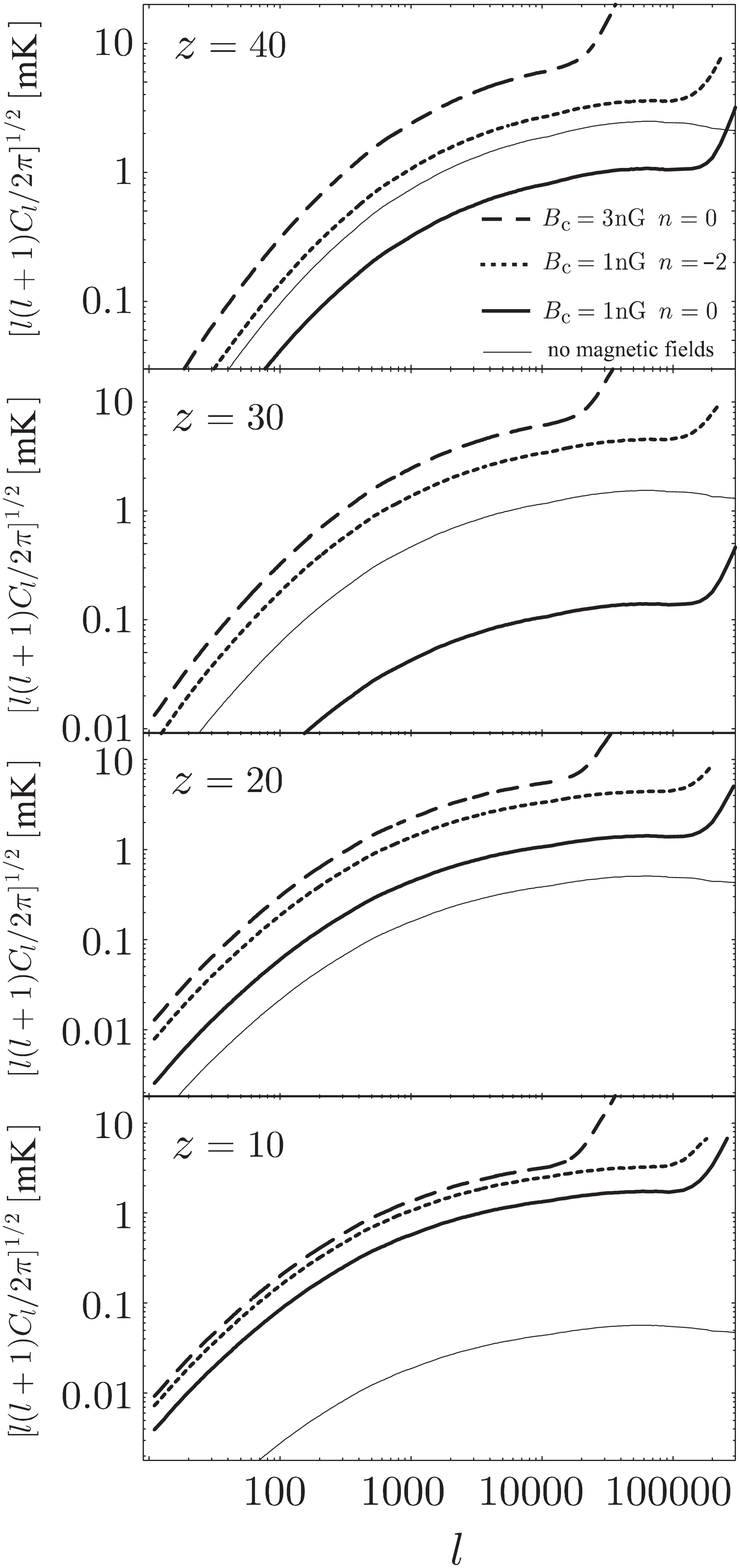}
  \end{center}
  \caption{The angler power spectra of CMB brightness temperature
  fluctuations by the 21cm line at the frequency $v_*/(1+z)$ for
  $z=40, 30, 20$, and $10$ from the top to bottom panels,
  respectively.  The dashed, dotted and solid lines represent the
  spectra for the models with $B_{\rm c}=3$nGauss and $n=0$, $B_{\rm c}=1$nGauss
  and $n=-2$ and $B_{\rm c}=1$nGauss and $n=0$, respectively.  The thin
  solid line in each panel is the spectrum of the model without the
  magnetic fields.  }
  \label{fig:llspectrum}
\end{figure}

Next, let us show the angler power spectra of the
brightness temperature fluctuations by the 21cm line for the various primordial
magnetic fields in Fig.~\ref{fig:llspectrum}.  For reference, we plot
the angler power spectra without magnetic fields, which are consistent
with previous works by \citet{bharadwaj-ali}.

The primordial magnetic fields affect the overall amplitude of the
angular power spectrum and the shape of the angular power spectrum on
large $l$'s.  The modification on the shape of the angular power
spectrum is caused by the density fluctuations generated by the
magnetic fields on small scales.  
From Eq.~(\ref{pert-brightness}),
the brightness temperature fluctuations are determined by the density 
fluctuations $\delta_{\rm m}$ and the spin temperature fluctuations 
$\delta T_{\rm s}/T_{\rm s}$.  
Applying the steady state approximation and $y \ll 1$ to
Eq.~(\ref{evo-deltaspin}), we can obtain $\delta T_{\rm s}/T_{\rm s}$ 
in the leading order of $y$ as 
\begin{equation}
{\delta T_{\rm s} \over T_{\rm s}} = y \left( {T_{\rm k}\over T_{\rm
\gamma}} -\gamma \right) \delta _{\rm m},
\end{equation}
where we ignore the temperature derivative term of $\kappa$ and
substitute $T_{\rm s}$ with Eq.~(\ref{approx-spin}).  
Accordingly Eq.~(\ref{pert-brightness}) leads to 
\begin{equation}
\delta T_{\rm b} 
= { \tau \over 1+z } y \left[ 2 T_{\rm k }- (1+\gamma)T_\gamma \right]
\delta_{\rm m}.
\label{eq:deltbapprox}
\end{equation}
It turns out that the brightness temperature fluctuations 
are simply proportional to the
density fluctuations.  
Therefore the additional ``blue'' density power spectrum
produced by the primordial magnetic fields on small scales appears as
the ``blue'' angler power spectrum on large $l$'s.  The magnetic Jeans
scale corresponds to $l_{\rm MJ} \sim 3\times 10^{5}(1 {\rm
nGauss}/B_{\rm c})$.  The angular power spectrum $C_l$ increases until
$l_{\rm MJ}$ and starts to show the oscillatory behavior above
$l>l_{\rm MJ}$.

The evolution of the power spectrum amplitude with the primordial
magnetic fields is classified into two regimes: $T_\gamma > T_{\rm s}$
and $T_\gamma < T_{\rm s}$.  In $T_\gamma > T_{\rm s}$, neutral
hydrogens absorb the CMB photons while neutral hydrogens emit the 21cm
line into CMB in $T_\gamma < T_{\rm s}$.  If there is no magnetic
fields, $T_\gamma > T_{\rm s}$ until the formation process of stars
and galaxies takes place.  On the other hand, the dissipation of the
primordial magnetic fields makes the hydrogen temperature higher than
the CMB temperature and $T_{\rm s}$ can exceed $T_\gamma$ at
high redshift as mentioned above.

In the regime of $T_\gamma > T_{\rm s}$, the amplitude of the angular
power spectrum with the primordial magnetic fields declines as is the
case with no magnetic fields due to the fact that $T_{\rm s}$
approaches to $T_\gamma$.  However, the decline of the power spectrum
is more prominent for the model with the primordial magnetic fields
(see bold and thin solid lines on the top and middle panels of
FIG.~\ref{fig:llspectrum}).  The reason is following.
For the model without magnetic fields, $T_\gamma \gg T_{\rm k}$.
From Eq.~(\ref{eq:deltbapprox}), therefore, 
the angular power spectrum $C_{l} \propto \vert
(1+\gamma)T_\gamma \vert^2$.  If the primordial magnetic fields exist,
on the other hand, $T_{\rm k}$ increases due to the dissipation.
Accordingly $C_{l} \propto \vert (1+\gamma)T_\gamma -2T_{\rm k}\vert^2
< \vert (1+\gamma)T_\gamma \vert^2$ as far as $T_{\rm k} 
< (1+\gamma)T_{\gamma}$. 
Therefore the suppression of the
angular power spectrum of the model with the primordial magnetic
fields is more prominent.

Once the hydrogen temperature $T_{\rm k}$ exceeds the CMB temperature
$T_\gamma$ (or equivalently $T_{\rm s} > T_\gamma$), the amplitude of
the brightness temperature fluctuations starts growing as the universe
evolves.  In the low redshift universe, the hydrogen temperature
becomes high enough that we can assume the CMB temperature is
negligible to the hydrogen temperature, $T_{\gamma} \ll T_{\rm k}$.
Hence the fluctuations of the brightness temperature can be
approximated as
\begin{equation}
\delta T_{\rm b} 
= { \tau  \over 1+z }  2y T_{\rm k } \delta_{\rm m}.
\end{equation}
Perhaps one might think that the amplitude of the magnetic fields 
can be determined by measuring the angular power spectrum since 
$T_{\rm k}$ depends on the amplitude.  However it is not the case.  
Because $y$ is related to the hydrogen temperature through $C_{10}/
T_{\rm k}$, the overall dependence of $\delta T_{\rm b}$ on the hydrogen
temperature comes from $C_{10}$.  In $\delta T_{\rm b} >100$K,
$C_{10}$ is not a strong function of the hydrogen temperature.
Accordingly, the amplitude of the angular spectrum does not depend much
on the strength of the primordial magnetic fields as is shown in
Fig.~\ref{fig:llspectrum}.  Therefore, in the regime
of $T_\gamma < T_{\rm s}$, we can only measure the strength of the primordial
magnetic fields through the turnover of the spectrum on high $l$'s.

\section{summary}

In this paper we investigate the effect of the primordial magnetic
fields on the CMB brightness temperature fluctuations produced by the
hydrogen 21cm line.

The brightness temperature fluctuations depend on the density
fluctuations, the hydrogen temperature and the ionization fraction whose
evolutions are affected by the primordial magnetic fields: the
dissipation of the magnetic field energy works as heat source and the
Lorentz force generates the density fluctuations after recombination.
The primordial magnetic fields with nano Gauss strength can heat 
hydrogens up to several thousands degrees Kelvin by the
ambipolar diffusion and produce dominant additional ``blue power'' in
the matter density spectrum on scales smaller than $100$kpc.

Through these effects, the primordial magnetic fields modify the
angular power spectrum of the brightness temperature fluctuations.
First, an additional blue spectrum can be found due to the existence
of the blue power in the matter density spectrum.  Secondly, the
overall amplitude of the angular spectrum can be modified by the
magnetic fields.  If the hydrogen temperature is lower than the
CMB temperature, the amplitude becomes smaller than the one
without magnetic fields.  The difference depends on the hydrogen
temperature.  On the other hand, if the hydrogen temperature becomes
higher than the CMB temperature, the amplitude of the angular
power spectrum, which depends little on the magnetic field amplitude
in this case, exceeds the one without magnetic fields.

Perhaps each feature of the brightness temperature fluctuations is not a direct evidence
for the existence of the primordial magnetic fields.  Possible
existence of the isocurvature mode on small scales can also boost 
the spectrum on small scales,  
while the higher hydrogen temperature
might be achieved by the other heat source, e.g., the decaying dark
matter.  However, both blue spectrum and modification of overall
amplitude of the angular power spectrum may provide a unique evidence
for the existence of the primordial magnetic fields.  Or at least we
can set a stringent constraint on the shape and the amplitude of the 
primordial magnetic fields.  

In this paper, we ignore the reionization process due to the ordinary 
astronomical objects, i.e., stars, which provides   
a significant impact on the brightness temperature fluctuations, since
we are focus on the epoch before reionization.  The reionization 
can be as late as $z\sim 9$ from the latest WMAP results \citep{wmap3}.  
Once the reionization process starts, 
the ionizing UV photons produced by stars make
the hydrogen temperature high and the Ly-$\alpha$ pumping of the
hydrogen 21cm transitions efficient.  Accordingly, the coupling
between the spin temperature and the hydrogen temperature becomes
stronger.  This reionization process could be also affected by 
the primordial magnetic fields due to the
modification of the matter power spectrum which induces the star
formation~\citep{tashiro-reion}.  Therefore we need to consider 
the reionization process with the primordial magnetic fields in order
to investigate the role of the magnetic fields below $z \la 10$, which
is beyond the scope of this paper.

\section*{Acknowledgements}
N.S. is supported by a Grant-in-Aid for Scientific Research from the
Japanese Ministry of Education (No. 17540276).

\end{document}